\documentstyle[a4,12pt,epsf]{article}
\newcommand{\be}{\begin{equation}}
\newcommand{\ee}{\end{equation}}
\newcommand{\egg}{$\eta \rightarrow \gamma \gamma \;$}
\newcommand{\epg}{$\eta \rightarrow \pi^0 \gamma \gamma \;$}
\newcommand{\sla}{\hspace{-0.5 em}/}
\newdimen\gap \gap=2pt
\newdimen\gaq \gaq=4pt
\newdimen\columnheight
\newdimen\columndepth
\newcommand{\col}[1]{\setbox255=\hbox{#1}
               \columnheight=\ht255 \columndepth=\dp255
               \advance\columnheight by \gaq
               \advance\columndepth by \gaq
               {\vrule height\columnheight depth\columndepth width0pt}#1}
\begin{document}
\baselineskip 1.5pc
\begin{center}
{\Large \bf $\eta$-Meson Decays and Strong $U_A(1)$ Breaking}
\vskip 5mm
{\Large \bf in the Three-Flavor Nambu-Jona-Lasinio Model}
\vskip 10mm
M. Takizawa\footnote{E-mail address: takizawa@ins.u-tokyo.ac.jp}\\
{\it Institute for Nuclear Study, University of Tokyo,} \\
{\it Tanashi, Tokyo 188, Japan}\\
\vskip 5mm
Y. Nemoto\footnote{E-mail address: nemoto@th.phys.titech.ac.jp} and 
M. Oka\footnote{E-mail address: oka@th.phys.titech.ac.jp}\\
{\it Department of Physics, Tokyo Institute of Technology,} \\
{\it Meguro, Tokyo 152, Japan} 
\end{center}
\vskip 5mm
\begin{abstract}
\baselineskip 1.5pc
    We study the \egg and \epg decays using an extended three-flavor 
Nambu-Jona-Lasinio model that includes the 't~Hooft instanton induced 
interaction.  
We find that the $\eta$-meson mass, the \egg decay width and the \epg 
decay width are in good agreement with the experimental values when the 
$U_{A}(1)$ breaking is strong and the flavor $SU(3)$ singlet-octet 
mixing angle $\theta$ is about zero.  
The effects of the $U_A(1)$ breaking on the baryon number one and two systems 
are also studied. 
\end{abstract}
\section{Introduction}
\hspace*{\parindent}It is well known that the QCD action has an approximate 
$U_L(3) \times U_R(3)$ chiral symmetry and its subsymmetry, $U_A(1)$
symmetry, is explicitly broken by the anomaly.  The $U_A(1)$ symmetry
breaking is manifested in the heavy mass of the $\eta'$ meson. 
\par
    The physics of the $\eta$ and $\eta'$ mesons have been extensively
studied in the $1/N_C$ expansion approach \cite{tHooft74}. In the $N_C
\to \infty$ limit, the $U_A(1)$ anomaly is turned off and then the
$\eta$ meson becomes degenerate with the pion and the $\eta'$ meson
becomes a pure $\bar ss$ state with $m_{\eta'}^2 (N_C \to \infty) = 2
m_K^2 - m_\pi^2 \simeq (687 \,{\rm MeV})^2$ \cite{Vene79}.  So the
$U_A(1)$ anomaly pushes up $m_{\eta}$ by about 400MeV and $m_{\eta'}$
by about 300MeV. It means that not only the $\eta'$ meson but also the
$\eta$ meson is largely affected by the $U_A(1)$ anomaly. 
\par
    In order to understand the role of the $U_A(1)$ anomaly in the
low-energy QCD, it may be important to study the $\eta$-meson decays
as well as its mass and decay constant.  Among the $\eta$-meson
decays, \egg and \epg decays are interesting.  
They have no final state interactions and involve only neutral mesons
so that the electromagnetic transitions are induced only by the
internal (quark) structure of the mesons.
The \egg decay is
related to the Adler-Bell-Jackiw triangle anomaly \cite{ABJ69} through
the partial conservation of axialvector current hypothesis.  For the
\epg decay, it is known that the chiral perturbation theory (ChPT) gives
too small prediction in the leading order and higher order terms are
expected to be dominant. 
\par
    The purpose of this paper is to study the \egg and \epg decays in the 
framework of the three-flavor Nambu-Jona-Lasinio (NJL) model as a chiral 
effective quark lagrangian of the low-energy QCD.
The three-flavor NJL model which involves the $U_L(3) \times U_R(3)$ symmetric 
four-quark interaction and the six-quark flavor-determinant interaction 
\cite{KKM71} incorporating effects of the $U_A(1)$ anomaly is used widely 
in recent years to study such topics as the quark condensates in vacuum, the 
spectrum of low-lying mesons, the flavor-mixing properties of the low-energy
hadrons, etc. \cite{HK94}.  In this approach the 
effects of the explicit breaking of the chiral symmetry by the current quark 
mass term and the $U_A(1)$ anomaly on the \egg and \epg decay amplitudes 
can be calculated consistently with those on the $\eta$-meson mass, $\eta$ 
decay constant and mixing angle within the model applicability.
\section{Three-flavor Nambu-Jona-Lasinio model}
\hspace*{\parindent}We work with the following NJL model lagrangian density:
\begin{eqnarray}
{\cal L} & = & {\cal L}_0 + {\cal L}_4 + {\cal L}_6 , \label{njl1} \\
{\cal L}_0 & = & \bar \psi \,\left( i \partial_\mu \gamma^\mu - \hat m 
\right) \, \psi \, ,
\label{njl2} \\
{\cal L}_4 & = & {G_S \over 2} \sum_{a=0}^8 \, \left[\, \left( 
\bar \psi \lambda^a \psi \right)^2 + \left( \bar \psi \lambda^a i \gamma_5 
\psi \right)^2 \, \right] \, ,
\label{njl3} \\
{\cal L}_6 & = & G_D \left\{ \, {\rm det} \left[ \bar \psi_i (1 - \gamma_5) 
\psi_j \right] + {\rm det}  \left[ \bar \psi_i (1 + \gamma_5) \psi_j 
\right] \, \right\} \, .
\label{njl4}
\end{eqnarray}
Here the quark field $\psi$ is a column vector in color, flavor and Dirac 
spaces and $\lambda^a (a=0\ldots 8)$ is the $U(3)$ generator in flavor space. 
The free Dirac lagrangian ${\cal L}_0$ incorporates the current quark mass 
matrix $\hat m = {\rm diag}(m_u, m_d, m_s)$ which breaks the chiral 
$U_L(3) \times U_R(3)$ invariance explicitly. ${\cal L}_4$ is a QCD 
motivated four-fermion interaction, which is chiral $U_L(3) \times U_R(3)$
invariant.  The 't Hooft determinant ${\cal L}_6$ represents the $U_A(1)$
anomaly.  It is a $3 \times 3$ determinant with respect to flavor with 
$i,j = {\rm u,d,s}$.    
\par
    Quark condensates and constituent quark masses are self-consistently
determined by the gap equations in the mean field approximation.
The covariant cutoff $\Lambda$ is introduced to regularize the
divergent integrals. The pseudoscalar channel quark-antiquark
scattering amplitudes are then calculated in the ladder approximation.
{}From the pole positions of the scattering amplitudes, the pseudoscalar
meson masses are determined. We define the effective meson-quark 
coupling constants $g_{\eta q q}$ and $g_{\pi q q}$ by introducing 
additional vertex lagrangians,
\be
  {\cal L}_{\eta q q} = g_{\eta q q} \overline{\psi} i\gamma_{5}\lambda^{\eta}
\psi \phi_{\eta},
\ee
\be
  {\cal L}_{\pi  q q} = g_{\pi  q q} \overline{\psi} i\gamma_{5}\lambda^{3}
\psi \phi_{\pi^{0}}, 
\ee
with $ \lambda^{\eta}=\cos \theta \lambda^{8}-\sin \theta \lambda^{0}$.
Here $\phi$ is an auxiliary meson field introduced for convenience and 
the effective meson-quark coupling constants are calculated from the residues 
of the $q \bar{q}$-scattering amplitudes at the corresponding meson poles.
Because of the $SU(3)$ symmetry breaking, the flavor $\lambda^{8}-\lambda^{0}$
components mix with each other. Thus we solve the coupled-channel $q \bar{q}$
scattering problem for the $\eta$ meson. The mixing angle $\theta$ is 
obtained by diagonalization of the $q \bar{q}$-scattering amplitude.
It should be noted that $\theta$ depends on $q^2$.  
At $q^2 = m_{\eta}^2$,  $\theta$ represents the mixing angle of the 
$\lambda^8$ and $\lambda^0$ components in the $\eta$-meson state.  
In the usual effective pseudoscalar meson lagrangian approaches, the 
$\eta$ and $\eta'$ mesons are analyzed using the $q^2$-independent 
$\eta$-$\eta'$ mixing angle.  Because of the $q^2$-dependence, $\theta$ cannot 
be interpreted as the $\eta$-$\eta'$ mixing angle.  The origin of the 
$q^2$-dependence is that the $\eta$ and $\eta'$ mesons have the internal 
quark structures. 
The meson decay constant $f_{M} \; (M = \pi, K, \eta)$ is determined by 
calculating the quark-antiquark one-loop graph. 
The explicit expressions are found in \cite{TO95}.
\section{\egg decay amplitude}
\hspace*{\parindent}The  $\pi^0, \eta \to \gamma \gamma$ decay amplitudes 
are given by 
\begin{equation}
\langle \gamma (k_1) \gamma (k_2) \vert M (q) \rangle = i (2 \pi)^4 
\delta^4 (k_1 + k_2 -q) \varepsilon_{\mu \nu \rho \sigma} 
\epsilon^{\mu}_1 \epsilon^{\nu}_2 k_1^{\rho} k_2^{\sigma} 
\widetilde{\cal T}_{M \to \gamma \gamma}(q^2) \, , \label{ampl}
\end{equation}
where $\epsilon_1$ and $\epsilon_2$ are the polarization vectors of the 
photon.  By calculating the pseudoscalar-vector-vector type quark triangle 
diagrams, we get the following results.
\begin{eqnarray}
\widetilde{\cal T}_{\pi^0 \to \gamma \gamma} & = & \frac{\alpha}{\pi} g_{\pi}
 F(u,\pi^0) \, , \label{amplpi}  \\ 
\widetilde{\cal T}_{\eta \to \gamma \gamma} & = & \frac{\alpha}{\pi} g_{\eta}
\frac{1}{3\sqrt{3}}  \Big[ \cos \theta \left\{ 5 F(u,\eta) - 2 F(s,\eta) 
\right\}\nonumber\\
& & \qquad \qquad 
- \sin \theta \sqrt{2} \left\{ 5 F(u,\eta) + F(s,\eta) \right\}
\Big] \, .  \label{ampleta}
\end{eqnarray} 
Here $\alpha$ is the fine structure constant of QED and $F(a,M)$ ($a=u,s$ and
$M=\pi^0,\eta$) is defined as
\begin{equation}
F(a,M) = \int_0^1 dx \int_0^1 dy \,
         \frac{2 (1-x) M_a}{M_a^2 - m_M^2 x (1-x) (1-y)} \, . \label{fam}
\end{equation}
Then the $M \to \gamma \gamma$ decay width $\Gamma(M \to \gamma \gamma)$ is 
given by
$\Gamma(M \to \gamma \gamma) = 
\vert \widetilde{\cal T}_{M \to \gamma \gamma} \vert^2 m_M^3 /(64 \pi)$.
\par
    In the chiral limit, the pion mass vanishes and $F(u,\pi^0)$ becomes 
$1/M_u$.  In this limit, the Goldberger-Treiman (GT) relation at the quark 
level, $M_u = g_{\pi} f_{\pi}$, holds in the NJL model and this leads to 
$\widetilde{\cal T}_{\pi^0 \to \gamma \gamma} = \alpha/(\pi f_{\pi})$ which 
is same as the tree-level results in the Wess-Zumino-Witten lagrangian 
approach \cite{WZW}.  It should be 
mentioned that we have to integrate out the triangle diagrams without 
introducing a cutoff $\Lambda$ in order to get the above result  though the 
cutoff is introduced in the gap equations in the NJL model.  
In the $U(3)_L \times U(3)_R$ version of the NJL model, 
the WZW term has been derived using the bosonization method with 
the heat-kernel expansion \cite{ER86,W89}.  In their approach, $O(1/\Lambda)$
term has been neglected and it is equivalent to taking the $\Lambda
\to \infty$ limit. 
\section{\epg decay amplitude}
\hspace*{\parindent}The \epg decay amplitude is given by
\be
  \langle \pi^{0}(p_{\pi}) \gamma(k_{1},\epsilon_{1}) \gamma(k_{2},
\epsilon_{2}) | \eta(p) \rangle = i (2\pi)^{4} \delta^{4}
(p_{\pi}+k_{1}+k_{2}-p) \epsilon_{1}^{\mu} \epsilon_{2}^{\nu} T_{\mu\nu}\, .
\ee
The dominant contributions to this process in this model are the quark-box 
diagrams. 
Following the evaluation of the quark-box diagrams performed in \cite{Ng93},
we obtain
\be
  T_{\mu\nu}=-i \frac{1}{\sqrt{3}}(\cos \theta -\sqrt{2} \sin \theta )
e^{2}g_{\eta q q}g_{\pi q q} \int \frac{d^{4}q}{(2\pi)^{4}} 
\sum_{i=1}^{6} U_{\mu\nu}^{i}, \label{epgg1}
\ee
with
\begin{eqnarray}
  U_{\mu\nu}^{1} & = & {\rm tr} \left\{ \gamma_{5} \frac{1}{q \sla 
- M + i\epsilon} \gamma_{5} \frac{1}{q \sla + p \sla - k \sla_{1}  
- k \sla_{2} - M +i\epsilon} \right. \nonumber \\
  & \times & \left. \gamma_{\nu} \frac{1}{q \sla + p \sla - k \sla_{1} 
- M + i\epsilon} \gamma_{\mu} \frac{1}{q \sla + p \sla - M +i\epsilon} 
\right\} , \label{umn1}\\
  U_{\mu\nu}^{2} & = & {\rm tr} \left\{ \gamma_{5} \frac{1}{q \sla 
- M + i\epsilon} \gamma_{5} \frac{1}{q \sla + k \sla_{2} - M +i\epsilon} 
\right. \nonumber \\
  & \times & \left. \gamma_{\nu} \frac{1}{q \sla + p \sla - k \sla_{1} 
- M + i\epsilon} \gamma_{\mu} \frac{1}{q \sla + p \sla - M +i\epsilon} 
\right\} , \\
  U_{\mu\nu}^{3} & = & {\rm tr} \left\{ \gamma_{5} \frac{1}{q \sla 
- M + i\epsilon} \gamma_{\nu} \frac{1}{q \sla + k \sla_{2} - M +i\epsilon} 
\right. \nonumber \\
  & \times & \left. \gamma_{\mu} \frac{1}{q \sla + k \sla_{1} + k \sla_{2} 
- M + i\epsilon} \gamma_{5} \frac{1}{q \sla + p \sla - M +i\epsilon} 
\right\} , \\
  U_{\mu\nu}^{4} & = & U_{\nu\mu}^{1}(k_{1} \leftrightarrow k_{2}), \\
  U_{\mu\nu}^{5} & = & U_{\nu\mu}^{2}(k_{1} \leftrightarrow k_{2}), \\
  U_{\mu\nu}^{6} & = & U_{\nu\mu}^{3}(k_{1} \leftrightarrow k_{2}). 
\label{umn2}
\end{eqnarray}
Here $M$ is the constituent u,d-quark mass. 
Because the loop integration in (\ref{epgg1}) is not divergent, 
we again do not use 
the UV cutoff. Then the gauge invariance is preserved. The inclusion of the
cutoff that is consistent with the gap equation will break the gauge 
invariance and make the present calculation too complicated.
Note that the strange quark does not contribute to the loop.
\par
   On the other hand the amplitude $T_{\mu\nu}$ has a general form required 
by the gauge invariance \cite{Eck88}
\begin{eqnarray}
  T^{\mu\nu} & = & A(x_{1},x_{2})(k_{1}^{\nu} k_{2}^{\mu} - k_{1} \cdot k_{2} 
g^{\mu\nu}) \nonumber \\
  & + & B(x_{1},x_{2}) \left[ -m_{\eta}^{2} x_{1}x_{2} g^{\mu\nu} -
\frac{k_{1} \cdot k_{2}}{m_{\eta}^{2}} p^{\mu}p^{\nu} + x_{1} k_{2}^{\mu} 
p^{\nu} + x_{2} p^{\mu} k_{1}^{\nu} \right],
\end{eqnarray}
with $x_{i}=p \cdot k_{i}/m_{\eta}^{2}$.
With $A$ and $B$, 
the differential decay rate with respect to the energies of the two photons 
is given by
\begin{eqnarray}
  \frac{d^{2} \Gamma}{dx_{1} dx_{2}} & = & \frac{m_{\eta}^{5}}{256 \pi^{2}} 
\left\{ \left|  A  + \frac{1}{2} B \right|^{2} \left[ 2(x_{1}+x_{2})
+\frac{m_{\pi}^{2}}{m_{\eta}^{2}} -1 \right]^{2} \right. \nonumber \\
  &  + &  \left. \frac{1}{4} \left| B \right| ^{2} \left[ 4 x_{1} x_{2} 
- \left[ 2(x_{1}+x_{2})+ \frac{m_{\pi}^{2}}{m_{\eta}^{2}}-1 \right] 
\right] ^{2} \right\}  \, . \label{gx1x2}
\end{eqnarray} 
Though the mass of $\eta$ as a $\bar{q} q$ bound state  depends on 
$G_{D}^{\rm eff}$,
we use the experimental value $m_{\eta}=547$ MeV in evaluating (\ref{gx1x2}).
The Dalitz boundary is given by two conditions:
\be
  \frac{1}{2} \left( 1-\frac{m_{\pi}^{2}}{m_{\eta}^{2}} \right) \leq  x_{1}
+x_{2} \leq 1- \frac{m_{\pi}}{m_{\eta}} ,
\ee
and
\be
  x_{1}+x_{2}-2 x_{1} x_{2} \leq \frac{1}{2} 
\left(1-\frac{m_{\pi}^{2}}{m_{\eta}^{2}} \right) .
\ee
In evaluating  (\ref{umn1})-(\ref{umn2}), one only has to identify the 
coefficients of 
$p^{\mu} p^{\nu}$ and $g^{\mu\nu}$. Details of the calculation are given 
in \cite{Ng93}. Defining ${\cal A}$ and ${\cal B}$ by
\be
  \int \frac{d^{4}q}{(2\pi)^{4}} \sum_{i=1}^{6} U_{i}^{\mu\nu} 
= -i\left( {\cal A} g^{\mu\nu} + {\cal B} \frac{p^{\mu} p^{\nu}}{m_{\eta}^{2}}
 + \cdots \right) ,
\ee
we find $A$ and $B$ as
\begin{eqnarray}
  A & = & \frac{1}{\sqrt{3}} (\cos \theta - \sqrt{2} \sin \theta) e^{2} 
g_{\pi q q} g_{\eta q q} \frac{2}{m_{\eta}^{2} \sigma} 
\left[ {\cal A}- 2 x_{1} x_{2} \
\frac{{\cal B}}{\sigma} \right] , \\
  B & = & \frac{1}{\sqrt{3}} (\cos \theta - \sqrt{2} \sin \theta) e^{2} 
g_{\pi q q} g_{\eta q q} \frac{2}{m_{\eta}^{2}} \frac{{\cal B}}{\sigma} ,
\end{eqnarray}
with
\be
  \sigma = \frac{(k_{1}+k_{2})^{2}}{m_{\eta}^{2}} = 2(x_{1}+x_{2})
+\frac{m_{\pi}^{2}}{m_{\eta}^{2}} -1 .
\ee
We evaluate  ${\cal A}$ and ${\cal B}$ numerically and further integrate 
(\ref{gx1x2}) to obtain the \epg decay rate. 
\section{Numerical Results}
\hspace*{\parindent}The recent experimental results of the 
$\pi^0, \eta \to \gamma \gamma$ decay
 widths are 
$\Gamma(\pi^0 \to \gamma \gamma) = 7.7 \pm 0.6 \, {\rm eV}$ and 
$\Gamma(\eta \to \gamma \gamma) = 0.510 \pm 0.026 \, {\rm keV}$ \cite{PDG} 
and the reduced amplitudes are 
\begin{eqnarray}
\left\vert \widetilde{\cal T}_{\pi^0 \to \gamma \gamma} \right\vert & = &
(2.5 \pm 0.1) \times 10^{-11} \, [{\rm eV}]^{-1} \, , \label{exppidw} \\
\left\vert \widetilde{\cal T}_{\eta \to \gamma \gamma} \right\vert & = &
(2.5 \pm 0.06) \times 10^{-11} \, [{\rm eV}]^{-1} \, . \label{expetadw}
\end{eqnarray}
{}From Eq. (\ref{amplpi}) and Eq. (\ref{ampleta}), we get 
$\widetilde{\cal T}_{\eta \to \gamma \gamma} = (5/3) 
\widetilde{\cal T}_{\pi^0 \to \gamma \gamma}$ in the $U_A(1)$ limit.  
Therefore in order to reproduce the experimental value of 
$\widetilde{\cal T}_{\eta \to \gamma \gamma}$, the effect of the 
$U_A(1)$ anomaly should reduce 
$\widetilde{\cal T}_{\eta \to \gamma \gamma}$ by a factor 3/5.
On the other hand, the experimental value of the \epg decay width 
is \cite{PDG}
\be
  \Gamma_{exp} \left( \eta \rightarrow \pi^{0} \gamma \gamma \right) 
= 0.85 \pm 0.19 {\rm \ eV}. \label{expepdw}
\ee
\par
    In our theoretical calculations, the parameters of the NJL model are
the current quark masses $m_{u}=m_{d},m_{s}$,
the four-quark coupling constant $G_{S}$, the six-quark determinant coupling
constant $G_{D}$ and the covariant cutoff $\Lambda$. We take $G_{D}$ as a 
free parameter and study $\eta$ meson properties as functions of $G_{D}$.
We use the light current quark masses $m_{u}=m_{d}=8.0$ MeV to reproduce 
$M_u = M_d \simeq 330$ MeV ($\simeq 1/3 M_N$) which is the value
usually used in the  
nonrelativistic quark model.
Other parameters, $m_{s},\; G_{D}$, and $\Lambda$, are determined
so as to reproduce the isospin averaged observed masses, $m_{\pi}, m_{K}$,
and the pion decay constant $f_{\pi}$.
\par
    We obtain $m_{s}=193$ MeV, $\Lambda=783$ MeV, $M_{u,d}=325$ MeV and 
$g_{\pi q q}=3.44$, which are almost independent of $G_{D}$.
The ratio of the current s-quark mass to the current u,d-quark mass is 
$m_s/m_u = 24.1$, which agrees well with $m_s/\hat{m} = 25 \pm 2.5$ 
 ($\hat{m} = \frac{1}{2} (m_u + m_d)$) derived from ChPT \cite{GL82}.
The kaon decay constant $f_K$ is the prediction and is almost independent of 
$G_D$.  We have obtained $f_K = 97$ MeV which is about 
14\% smaller than the observed value.  We consider this is the typical 
predictive power of the NJL model in the strangeness sector.
\par
    Table 1 summarizes the fitted results of the model parameters and the 
quantities necessary for calculating the \egg and \epg decay widths 
which depend on $G_{D}$.  
We define dimensionless parameters 
$G_{D}^{\rm eff} \equiv - G_{D} (\Lambda / 2 \pi)^{4} \Lambda N_{c}^{2}$ and 
$G_{S}^{\rm eff} \equiv G_{S} (\Lambda / 2 \pi)^{2} N_{c}$. 
When $G_{D}^{\rm eff}$ is zero, our lagrangian does not cause the flavor 
mixing and therefore the ideal mixing is achieved. 
The ``$\eta$'' is purely $u\bar u + d\bar d$ and is degenerate 
to the pion in this limit.
\begin{table}[t]
\caption{The parameters of the model, the \egg decay amplitude $
\widetilde{\cal T}_{\eta \to \gamma \gamma}$ and \epg decay width 
$\Gamma(\eta \to \pi^0 \gamma \gamma)$ for each $G_{D}^{\rm eff}$}
\begin{center}
\begin{tabular}{|c|c|c|c|c|c|c|c|c|} 
\hline
\col{$G_{D}^{\rm eff}$} & $G_{S}^{\rm eff}$ & $M_s$ [MeV] & $m_{\eta}$ [MeV] 
& $\theta$ [deg] & $g_{\eta qq}$ & 
$\widetilde{\cal T}_{\eta \to \gamma \gamma}$ [1/eV] & 
$\Gamma(\eta \to \pi^0 \gamma \gamma)$ [eV] \\ \hline
0.00 & 0.73 & 556 & 138.1 & -54.74 & 3.44 & 4.17$\times 10^{-11}$ & 2.88 \\
0.10 & 0.70 & 552 & 285.3 & -44.61 & 3.23 & 3.95$\times 10^{-11}$ & 2.46 \\
0.20 & 0.66 & 549 & 366.1 & -33.52 & 3.12 & 3.68$\times 10^{-11}$ & 2.06 \\
0.30 & 0.63 & 545 & 419.1 & -23.24 & 3.11 & 3.39$\times 10^{-11}$ & 1.71 \\
0.40 & 0.60 & 541 & 455.0 & -14.98 & 3.15 & 3.10$\times 10^{-11}$ & 1.42 \\
0.50 & 0.57 & 537 & 479.7 & -8.86  & 3.20 & 2.86$\times 10^{-11}$ & 1.20 \\
0.60 & 0.54 & 533 & 497.3 & -4.44  & 3.25 & 2.65$\times 10^{-11}$ & 1.04 \\
0.70 & 0.51 & 529 & 510.0 & -1.25  & 3.28 & 2.48$\times 10^{-11}$ & 0.92 \\
0.80 & 0.47 & 525 & 519.6 &  1.09  & 3.30 & 2.35$\times 10^{-11}$ & 0.84 \\
0.90 & 0.44 & 522 & 527.0 &  2.84  & 3.31 & 2.23$\times 10^{-11}$ & 0.77 \\
1.00 & 0.41 & 518 & 532.8 &  4.17  & 3.32 & 2.14$\times 10^{-11}$ & 0.71 \\
1.10 & 0.40 & 514 & 537.5 &  5.21  & 3.32 & 2.07$\times 10^{-11}$ & 0.67 \\
1.20 & 0.35 & 511 & 541.3 &  6.02  & 3.31 & 2.00$\times 10^{-11}$ & 0.63 \\
1.30 & 0.32 & 507 & 544.5 &  6.66  & 3.30 & 1.95$\times 10^{-11}$ & 0.61 \\
1.40 & 0.29 & 504 & 547.2 &  7.17  & 3.29 & 1.91$\times 10^{-11}$ & 0.58 \\
1.50 & 0.25 & 500 & 549.4 &  7.57  & 3.28 & 1.86$\times 10^{-11}$ & 0.56 \\
1.60 & 0.22 & 497 & 551.4 &  7.90  & 3.26 & 1.83$\times 10^{-11}$ & 0.55 \\
\hline
\end{tabular} 
\end{center}
\end{table}
\par
    We next discuss the $\pi^0 \to \gamma \gamma$ decay.  
The calculated result is 
$\widetilde{\cal T}_{\pi^0 \to \gamma \gamma} = 
2.50 \times 10^{-11} (1/{\rm eV})$
which agrees well with the observed value given in Eq. (\ref{exppidw}).
The current algebra result is 
$\widetilde{\cal T}_{\pi^0 \to \gamma \gamma} = 
\alpha/(\pi f_\pi) = 2.514 \times 10^{-11} (1/{\rm eV})$, 
and thus the soft pion limit is a good approximation for 
$\pi^0 \to \gamma \gamma$ decay.  
The chiral symmetry breaking affects
$\widetilde{\cal T}_{\pi^0 \to \gamma \gamma}$ in two ways.  One is
the deviation from  
the G-T relation and another is the matrix element of the triangle diagram 
$F(u,\pi^0)$.  Our numerical results are $g_\pi = 3.44$, $M_u/f_\pi = 3.52$
and $F(u,\pi^0) M_u = 1.015$, therefore the deviations from the soft pion 
limit are very small both in the G-T relation and the matrix element of the
triangle diagram.
\par
    Let us now turn to the discussion of the \egg decay.  The calculated 
results of the \egg decay amplitude 
$\widetilde{\cal T}_{\eta \to \gamma \gamma}$ are given in Table 1 and shown 
in Fig. 1. 
\begin{figure}[t]
\caption{Dependence of the $\eta \to \gamma \gamma $ decay amplitude
on the dimension-less coupling constant $G_{D}^{\rm eff}$. The horizontal
dashed line indicates the experimental value.\hfil\break} 
{\Large \mbox{$\widetilde{\cal T}_{\eta \to \gamma \gamma}$} [1/eV]} \\
\epsfxsize=380pt
\epsfbox{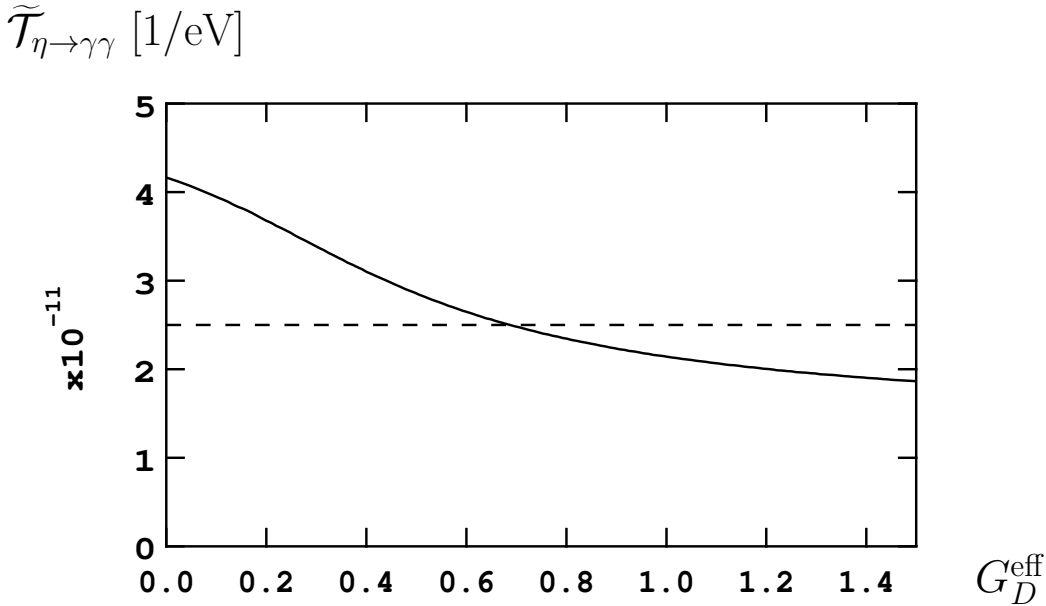}
\begin{flushright}
  \vspace*{-25mm} 
  {\Large $G_{D}^{\rm eff}$ \hspace*{20mm}} 
\end{flushright}
\end{figure}
The experimental value of the \egg decay amplitude is reproduced at about 
$G_D^{\rm eff} = 0.7$. The calculated $\eta$-meson mass at 
$G_D^{\rm eff} = 0.7$ is $m_{\eta} = 510$ MeV which is 7\% smaller than 
the observed mass.  $G_D^{\rm eff} = 0.7$ corresponds to 
$G_{D} \langle \overline{s} s \rangle / G_{S} =0.44$, suggesting that 
the contribution from ${\cal L}_{6}$ to the dynamical mass of 
the up and down quarks is 44\% of that from ${\cal L}_{4}$.
\par
    The mixing angle at $G_{D}^{\rm eff}=0.7$ is $\theta = -1.3^{\circ}$ and 
that indicates a strong OZI violation and a large (u,d)-s mixing. 
This disagrees with the ``standard'' value $\theta \simeq -20^{\circ}$ 
obtained in ChPT\cite{DHL88}. This is due to the 
stronger $U_{A}(1)$ breaking in the present calculation.
The difference mainly comes from the fact that the mixing angle 
in the NJL model depends on $q^{2}$ of the $\overline{q}q$ state 
and thus reflects the internal structure of the $\eta$ meson. 
On the contrary the analyses of ChPT\cite{DHL88} 
assume an energy-independent mixing angle, i.e., 
$\theta(q^2=m_{\eta}^{2})=\theta(q^2=m_{\eta'}^{2})$.
\par
    The $\eta$ decay constant is almost independent of $G_D$:
$f_{\eta} = 91.2$ MeV ($\simeq f_{\pi}$) at $G_{D}^{\rm
eff}=0.7$. Therefore it seems that the $\eta$ meson does not lose the
Nambu-Goldstone boson nature though its mass and mixing angle are
strongly affected by the $U_A(1)$ breaking interaction. 
\par
    The \epg decay gives us an independent information on the 
structure of the $\eta$ meson.  The calculated  
\epg decay widths are also given in Table 1 and shown in Fig. 2. 
\begin{figure}[t]
\caption{Dependence of the $\eta \to \pi^{0} \gamma \gamma $ decay
width on the dimension-less coupling constant $G_{D}^{\rm eff}$. The horizontal
solid line indicates the experimental value and the dashed
lines indicate its error widths. \hfil\break}
{\Large \mbox{\boldmath$\Gamma(\eta \to \pi^0 \gamma \gamma) $} [eV]} \\
\epsfxsize=360pt
\epsfbox{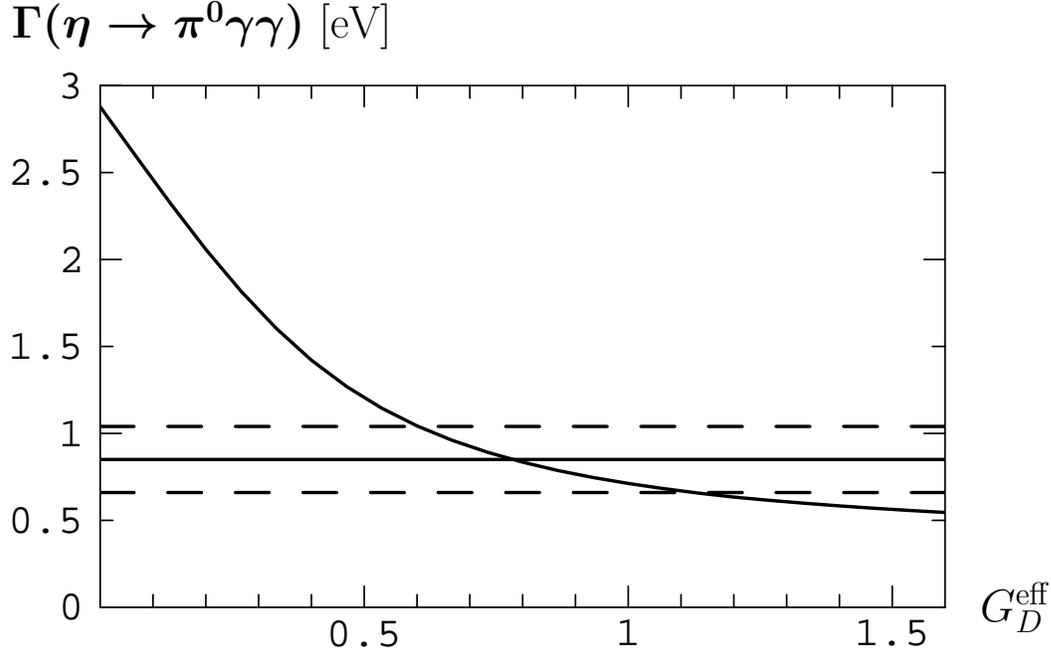} 
\begin{flushright}
  \vspace*{-15mm} 
  {\Large $G_{D}^{\rm eff}$ \hspace*{20mm}}  
\end{flushright}
\end{figure} 
At $G_{D}^{\rm eff}=0.70$, we obtain 
$\Gamma(\eta \rightarrow \pi^{0} \gamma \gamma )=0.92$, which is in good 
agreement with the experimental data shown in (\ref{expepdw}). 
This process was studied in ChPT\cite{ABBC92} and in the extended NJL model
\cite{BB95}.  The difference between these approaches and ours are discussed 
in \cite{NOT96}. 
\section{Effects of the $U_A(1)$ Anomaly in Baryons}
\hspace*{\parindent}Since the effects of the $U_A(1)$ anomaly are rather 
large in the pseudoscalar sector, it is natural to ask if one can see some 
effects in the baryon sector.  It was pointed out in \cite{SR89} that the 
instanton can play an important role in the description of spin-spin forces, 
particularly for light baryons.  The pattern of these effects can be very hard 
to disentangle from one-gluon exchange. The effects of the instanton induced 
interaction in baryon number $B=2$ systems were studied in \cite{OT89}.
It was shown that an attraction between two nucleons is obtained by
the two-body 
instanton induced interaction, while the three-body interaction is strongly 
repulsive in the H-dibaryon channel and makes the H-dibaryon almost unbound.
\par
   We estimate the effects of the $U_A(1)$ anomaly on the $B = 1$ and 
$B = 2$ systems by employing the six-quark determinant interaction given in 
Eq. (\ref{njl3}) whose strength was determined so as to reproduce the 
observed $\eta$-meson mass, the \egg decay width and the \epg decay width, 
namely, $G_D^{\rm eff} = 0.7$.  It is done by calculating the matrix 
elements of the the $U_A(1)$ breaking interaction hamiltonian with respect to 
unperturbed states of the MIT bag model and the nonrelativistic quark model 
(NRQM).  For $B = 2$ systems, we only consider the $(0S)^6$ configuration 
of the six valence quark states. Therefore, the matrix element with respect 
to such a state gives a measure of the contribution of the $U_A(1)$ breaking 
interaction either to the dibaryon or to the short-range part of the 
interaction between two baryons.  The determinant interaction induces not 
only three-body but also two-body interactions of valence quarks when the 
vacuum has a nonvanishing quark condensate.  The details of the 
calculation are described in \cite{MT93}.  
\par
Table 2 shows the contribution of 
the two-body term for $B = 1$. 
The contribution to the decuplet baryons vanishes in the $SU(3)$ limit 
and therefore comes only from the $SU(3)$ asymmetry of the quark wave
function. 
The three-body term does not contribute to the $B = 1$ states. 
Thus the  N$\Delta$ mass difference due to the $U_A(1)$ 
breaking interaction is about 15\% of the observed one.   
\begin{table}[t]
\caption{Contribution of the two-body term to octet and decuplet baryons. 
All the entries are in units of MeV}
\begin{center}
\begin{tabular}{|c|cccc|cccc|}
\hline
wave function & \col{$N$} & $\Sigma$ & $\Xi$ & $\Lambda$ 
& $\Delta$ & ${\Sigma ^*}$ & ${\Xi^*}$ & $\Omega$ \\
\hline
\col{MIT} & $-43.9$ & $-41.2.$ & $-41.2$ & $-42.9$ & $0$ &
$0.12$ & $0.12$ & $0$ \\
\col{NRQM} & $-40.88$ & $-36.6$ & $-36.6$ & $-39.4$ & $0$ &
$0.07$ & $0.07$ & $0$ \\
\hline
\end{tabular}
\end{center}
\end{table}
\par
    We next discuss the case of $B = 2$.  We consider all the possible 
channels which are made of two octet baryons listed in Table 3.
\begin{table}[t]
\caption{Baryon component, $SU(3)$ multiplet, spin, isospin and strangeness 
of the eight channels of two octet baryons}
\begin{center}
\begin{tabular}{|c|ccccc|}
\hline
channel & \col{Baryon component} & $SU(3)$ multiplet & Spin & Isospin & 
Strangeness \\
\hline
\col{I} & $NN$ & {\bf 10*} & 1 & 0 & 0 \\
\col{II} & $NN$ & {\bf 27} & 0 & 1 & 0 \\
\hline
\col{III} & $N\Sigma$ & {\bf 27} & 0 & 3/2 & -1 \\
\col{IV} & $N\Sigma-N\Lambda$ & {\bf 27} & 0 & 1/2 & -1 \\
\col{V} & $N\Sigma-N\Lambda$ & {\bf 10*} & 1 & 1/2 & -1 \\
\col{VI} & $N\Sigma$ & {\bf 10} & 1 & 3/2 & -1 \\
\col{VII} & $N\Sigma-N\Lambda$ & {\bf 8} & 1 & 1/2 & -1 \\
\hline
\col{VIII} & $H$ & {\bf 1} & 0 & 0 & -2 \\
\hline
\end{tabular}
\end{center}
\end{table}
Table 4 shows the contribution of the two-body term.  The channel VIII gets 
the strongest attraction, about $170$ MeV, and the channel VII gets the second 
strongest attraction.  The contributions of the three-body term to the 
H-dibaryon and strangeness $-1$ channels are given in Table 5.
It should be noted that the three-body term has no effect on the NN
channels, and that the contributions to the channels III, IV and V
reflect the $SU(3)$ breaking in the quark wave function.
The contributions of the three-body term in channels VI, VII and VIII are 
remarkable and one will be able to observe some effects experimentally.  

\begin{table}[t]
\caption{Contributions of the two-body term to the eight channels of two 
octet baryons listed in Table 3.  All the entries are in units of MeV}
\begin{center}
\begin{tabular}{|c|cc|ccccc|c|}
\hline
wave function & \col{I} &  II & III & IV & V & VI & VII & VIII \\
\hline
\col{MIT} & $-89.9$ & $-85.1$ & $-88.9$ & $-86.3$ & $-93.9$ & $-96.5$
& $-121.5$ & $-162.6$ \\
\col{NRQM} & $-120.2$ & $-105.3$ & $-102.3$ & $-103.9$ & $-118.3$ & $-116.7$
& $-148.0$ & $-182.6$ \\
\hline
\end{tabular}
\end{center}
\end{table}
\begin{table}[t]
\caption{Contributions of the three-body term to the H-dibaryon (VIII) and 
strangeness $-1$ two octet baryon channels (III-VII).  All the entries are 
in units of MeV}
\begin{center}
\begin{tabular}{|c|ccccc|c|}
\hline
wave function & \col{III} & IV & V & VI & VII & VIII \\
\hline
\col{MIT} & $-6\times 10^{-2}$ & $-6\times 10^{-2}$ &
$-7\times 10^{-2}$ & $20.7$ & $25.1$ & $40.7$ \\
\col{NRQM} & $-5\times 10^{-2}$ & $-5\times 10^{-2}$ &
$-5\times 10^{-2}$ & $28.3$ & $34.9$ & $56.1$ \\
\hline
\end{tabular}
\end{center}
\end{table}
\par
    We should comment on the difference between the determinant 
interaction  used here and the instanton-induced interaction 
used in ref. \cite{OT89}.  The relative contributions of the $U_A(1)$ 
breaking interaction within the baryonic sector or within the mesonic 
sector are similar for the two interactions. However, the ratio of
those in the baryonic sector to those in the mesonic sector is about
$\frac{4}{7}$.  
Namely, if one fixes the strength of the interaction so as to give the 
same mass difference of $\eta$ and $\eta'$, the effects of the
instanton-induced interaction in the baryonic sector would be about
$\frac{7}{4}$ stronger than those of the determinant interaction.  
After this correction the strength of the present $U_A(1)$ breaking
interaction is consistent with that used in the calculation of the
baryon-baryon interaction in ref. \cite{OT89}.
\section{Summary}
\hspace*{\parindent}We have studied the \epg  decay in the three-flavor NJL 
model that includes the $U_{A}(1)$ breaking six-quark determinant interaction.
The $\eta$ meson mass, the $\eta \to \gamma \gamma$ decay width and the \epg
decay width are reproduced well with a rather strong $U_{A}(1)$ breaking
interaction, that makes $\eta_{1}-\eta_{8}$ mixing angle $\theta 
\simeq 0^{\circ}$.  The effects of the $U_A(1)$ breaking on the baryon 
number one and two systems have been also studied.
\par
    Finally, we should note that the NJL model does not confine quarks.  
Since the NG bosons, $\pi$, K, and $\eta$, are strongly bound, the NJL
model can describe their properties fairly well.  However the $\eta'$ meson
state in the NJL model has an unphysical decay of the $\eta' \to q \bar q$.
Therefore we do not apply our model to the $\eta'$ meson.  Further study of 
the $U_A(1)$ breaking will require the construction of a framework which
can be applied to the $\eta'$ meson.
\pagebreak
\end{document}